\documentclass[twocolumn, twocolappendix]{aastex7}

\usepackage{amsmath}
\usepackage{color}

\begin{document}

\title{Dust-obscured Galaxies with Broken Power-law Spectral Energy Distributions Discovered by UNIONS}

%% argument which is the 16 digit ORCID. The syntax is:
%% \author[xxxx-xxxx-xxxx-xxxx]{Author Name}
%%
%% This will hyperlink the author name to the author's ORCID page. Note that
%% during compilation, LaTeX will do some limited checking of the format of
%% the ID to make sure it is valid. If the "orcid-ID.png" image file is 
%% present or in the LaTeX pathway, the OrcID icon will appear next to
%% the authors name.
%%

%\correspondingauthor{August Muench}
%\email{greg.schwarz@aas.org, gus.muench@aas.org}

\author[0009-0001-9947-6732]{Taketo Yoshida}
\affiliation{Graduate School of Science and Engineering, Ehime University, 2-5 Bunkyo-cho, Matsuyama, Ehime 790-8577, Japan}
\email[show]{yoshida@cosmos.phys.sci.ehime-u.ac.jp}

\author[0000-0002-7402-5441]{Tohru Nagao}
\affiliation{Research Center for Space and Cosmic Evolution, Ehime University, 2-5 Bunkyo-cho, Matsuyama, Ehime 790-8577, Japan}
\affiliation{Amanogawa Galaxy Astronomy Research Center, Kagoshima University, 1-21-35 Korimoto, Kagoshima 890-0065, Japan}
\email{tohru@cosmos.ehime-u.ac.jp}

\author[0000-0002-3531-7863]{Yoshiki Toba}
\affiliation{Department of Physical Sciences, Ritsumeikan University, 1-1-1, 
Noji-higashi, Kusatsu, Shiga 525-8577, Japan}
\affiliation{National Astronomical Observatory of Japan, 2-21-1 Osawa, Mitaka, Tokyo 181-8588, Japan}
\affiliation{Academia Sinica Institute of Astronomy and Astrophysics, 11F of Astronomy-Mathematics Building, AS/NTU, No.1, Section 4, Roosevelt Road, Taipei 10617, Taiwan}
\affiliation{Research Center for Space and Cosmic Evolution, Ehime University, 2-5 Bunkyo-cho, Matsuyama, Ehime 790-8577, Japan}
\email{toba@fc.ritsumei.ac.jp}

\author{Akatoki Noboriguchi}
\affiliation{Center for General Education, Shinshu University, 3-1-1 Asahi, Matsumoto, Nagano 390-8621, Japan}
\email{noboriguchi.astro@gmail.com}

\author{Kohei Ichikawa}
\affiliation{Waseda Research Institute for Science and Engineering, Faculty of Science and Engineering, Waseda University, 3-4-1 Okubo, Shinjuku, Tokyo 169-8555, Japan}
\email{ichikawa.waseda@gmail.com}

\author{Hendrik Hildebrandt}
\affiliation{Ruhr University Bochum, Faculty of Physics and Astronomy, Astronomical Institute (AIRUB), German Centre for Cosmological Lensing, 44780 Bochum, Germany}
\email{hendrik@astro.ruhr-uni-bochum.de}

\author{Naomichi Yutani}
\affiliation{Graduate School of Science and Engineering, Kagoshima University, 1-21-35 Korimoto, Kagoshima 890-0065, Japan}
\email{yutaninm@gmail.com}

\author{Kenneth C. Chambers}
\affiliation{Institute of Astronomy, University of Hawaii, 2680 Wood- lawn Drive, Honolulu, Hawaii 96822, USA}
\email{chambers@ifa.hawaii.edu}

\author[0009-0005-8071-682X]{Ryo Iwamoto}
\affiliation{Graduate School of Science and Engineering, Ehime University, 2-5 Bunkyo-cho, Matsuyama, Ehime 790-8577, Japan}
\email{iwamoto@cosmos.phys.sci.ehime-u.ac.jp}

\author{Seira Kobayashi}
\affiliation{Graduate School of Science and Engineering, Ehime University, 2-5 Bunkyo-cho, Matsuyama, Ehime 790-8577, Japan}
\email{seira@cosmos.phys.sci.ehime-u.ac.jp}

\author[0000-0003-3484-399X]{Masamune Oguri}
\affiliation{Center for Frontier Science, Chiba University, 1-33 Yayoi-cho, Inage-ku, Chiba 263-8522, Japan}
\affiliation{Department of Physics, Graduate School of Science, Chiba University, 1-33 Yayoi-Cho, Inage-Ku, Chiba 263-8522, Japan}
\email{masamune.oguri@chiba-u.jp}

\author[0000-0002-7934-2569]{Ken Osato}
\affiliation{Center for Frontier Science, Chiba University, 1-33 Yayoi-cho, Inage-ku, Chiba 263-8522, Japan}
\affiliation{Department of Physics, Graduate School of Science, Chiba University, 1-33 Yayoi-Cho, Inage-Ku, Chiba 263-8522, Japan}
\affiliation{Kavli Institute for the Physics and Mathematics of the Universe, Todai Institutes for Advanced Study, The University of Tokyo, 5-1-5 Kashiwanoha, Kashiwa, Chiba 277-8583, Japan}
\email{ken.osato@chiba-u.jp}

\author{Kohei Shibata}
\affiliation{Graduate School of Science and Engineering, Ehime University, 2-5 Bunkyo-cho, Matsuyama, Ehime 790-8577, Japan}
\email{shibata@cosmos.phys.sci.ehime-u.ac.jp}

\author[0009-0001-3910-2288]{Yuxing Zhong}
\affiliation{Faculty of Science and Engineering, Waseda University, 3-4-1 Okubo, Shinjuku, Tokyo 169-8555, Japan}
\email{yuxing.zhong.astro@gmail.com}

%\collaboration{20}{(AAS Journals Data Editors)}


\begin{abstract}

We report on the spectral energy distributions (SEDs) of infrared-bright dust-obscured galaxies (DOGs) with $(i - [22])_{\rm AB} \geq 7.0$.
Using photometry from the deep and wide Ultraviolet Near-Infrared Optical Northern Survey, combined with near-IR and mid-IR data from the UKIRT Infrared Deep Sky Survey and the Wide-field Infrared Survey Explorer, we successfully identified 382 DOGs in $\sim$ 170 deg$^2$. 
Among them, the vast majority (376 DOGs) were classified into two subclasses: bump DOGs (132/376) and power-law (PL) DOGs (244/376), which are dominated by star formation and active galactic nucleus (AGN), respectively.
Through the SED analysis, we found that roughly half (120/244) of the PL DOGs show ``broken'' power-law SEDs. 
The significant red slope from optical to near-IR in the SEDs of these ``broken power-law DOGs'' (BPL DOGs) probably reflects their large amount of dust extinction. 
In other words, BPL DOGs are more heavily obscured AGNs, compared to PL DOGs with non-broken power-law SEDs. 

\end{abstract}



\section{Introduction} \label{sec:intro}

Past observational studies have shown that the masses of supermassive black holes (SMBHs) and those of their host galaxies are strongly correlated, suggesting that they have co-evolved in each other (e.g., Magorrian et al. \citeyear{Magorrian98}; Marconi \& Hunt \citeyear{Marconi03}; Kormendy \& Ho \citeyear{Kormendy13}).
One plausible model to explain the physics of the co-evolution is a scenario considering major mergers of gas-rich galaxies (so-called wet mergers; e.g., Sanders et al. \citeyear{Sanders88}; Hopkins et al. \citeyear{Hopkins08}).
In this scenario, the merging/interaction of gas-rich galaxies induces a burst of
star-formation (SF) enshrouded by a large amount of dust. 
Then the gas losing its angular momentum falls onto the nucleus (e.g., Barnes and Hermquist \citeyear{Barnes91}), which turns on the active galactic nucleus (AGN) at the center of the dusty galaxy. 
The powerful radiation and/or outflow from the AGN blows out the surrounding gas and dust, making the system optically thin, which corresponds to an optically bright quasar.
Though this major merger scenario for the co-evolution has been widely discussed theoretically, its validity has not been fully tested observationally (c.f., Chiaberge et al. \citeyear{Chiaberge15}; Pierce et al. \citeyear{Pierce23}; Zhong et al. \citeyear{Zhong24}). 
This is mainly because the active SF and AGN phases are obscured by a large amount of dust (Blecha et al. \citeyear{Blecha18}), making it difficult to determine the evolutionary stage of each object in this scenario.

To study dusty populations of galaxies and AGNs, we focus on dust-obscured galaxies (DOGs: Dey et al. \citeyear{Dey08}). 
They are selected based on their extremely red color in optical to mid-infrared (MIR) wavelengths in the observed frame, which is caused by extreme dust extinction in optical and re-emission in infrared (IR). 
DOGs have been divided into two subclasses based on their optical-MIR  spectral energy distributions (SEDs), which are known as “bump DOGs” and “power-law (PL) DOGs” (e.g., Dey et al. \citeyear{Dey08}; Melbourne et al. \citeyear{Melbourne12}; Toba et al. \citeyear{Toba15}; Suleiman et al. \citeyear{Suleiman22}; Yu et al. \citeyear{Yu24}).
Bump DOGs are characterized by a bump feature (corresponding to rest-frame 1.6 $\mu$m bump) around near-infrared (NIR) in their SEDs mainly caused by the stellar photospheres of cool stars, which suggests that star-forming activity dominates those DOGs (e.g., Desai et al. \citeyear{Desai09}; Melbourne et al. \citeyear{Melbourne12}; Yu et al. \citeyear{Yu24}; see also Pope et al. \citeyear{Pope08}; Riguccini et al. \citeyear{Riguccini11}). 
PL DOGs are characterized by power-law shapes of their SEDs from optical to MIR wavelengths, and represent dusty AGNs (e.g., Bussmann et al. \citeyear{Bussmann09a}; Melbourne et al. \citeyear{Melbourne12}; Zou et al. \citeyear{Zou20}; Yu et al. \citeyear{Yu24}; see also Alonso-Herrero et al. \citeyear{Alonso-Herrero06}; Imanishi et al. \citeyear{Imanishi10}; Ichikawa et al. \citeyear{Ichikawa14}).
The presence of obscured AGNs in PL DOGs is also supported by X-ray studies, showing that the majority of PL DOGs have a large column density ($N_{\mathrm{H}} > 10^{22}$ cm$^{-2}$; e.g., Lanzuisi et al. \citeyear{Lanzuisi09}; Corral et al. \citeyear{Corral16}; Riguccini et al. \citeyear{Riguccini19}; Toba et al. \citeyear{Toba20}; Zou et al. \citeyear{Zou20}; Kayal \& Singh \citeyear{Kayal24}; Noboriguchi et al. \citeyear{Nobo25}).

Additionally, some DOGs are characterized by hotter dust ($\gtrsim$ 60K; Wu et al. \citeyear{Wu12}; Fan et al. \citeyear{Fan17}) compared to those of classical DOGs ($<$ 60K; Melbourne et al. \citeyear{Melbourne12}), which are known as Hot DOGs (Eisenhardt et al. \citeyear{Eisenhardt12}; Wu et al. \citeyear{Wu12}).
They are more luminous and more heavily obscured DOGs, compared with classical DOGs (e.g., Tsai et al. \citeyear{Tsai15}; Toba et al. \citeyear{Toba18}, \citeyear{Toba20}; Li et al. \citeyear{Li24}).

James Webb Space Telescope (JWST) has revealed a new dusty populations of galaxies in the early Universe. 
In particular, spatially compact and extremely red objects have been newly recognized, such as extremely red objects (EROs: Barro et al. \citeyear{Barro24}) or little red dots (LRDs: Matthee et al. \citeyear{Matthee24}), though their nature is currently unclear. 
Recently, a possible relationship between DOGs and these dusty populations was discussed based on similarity of their SEDs by Noboriguchi et al. (\citeyear{Nobo23}). 

Numerical simulations have suggested that gas-rich major mergers of galaxies experience a DOG phase around the peak of their star formation rate (Narayanan et al. \citeyear{Narayanan10}; Yutani et al. \citeyear{Yutani22}), suggesting that DOGs presumably trace the most active phase of gas-rich major mergers.
In other words, DOGs are considered a key population for understanding the co-evolutionary scenario of SMBHs and galaxies.

As described above, both observations and simulations suggest that DOGs evolve from the bump to PL phase (e.g., Dey \& NDWFS/MIPS Collaboration \citeyear{Dey09}; Bussmann et al. \citeyear{Bussmann12}; Yutani et al. \citeyear{Yutani22}).
This indicates a possible link with the merger-driven galaxy-SMBH co-evolution framework.

In summary, researching DOGs and their subclasses is effective in understanding the evolutionary stage of major mergers of gas-rich galaxies.
Constructing a large sample of DOGs is challenging because they are spatially rare and significantly faint in optical wavelengths, requiring a wide survey area and deep sensitivity.
Previous studies have identified up to several thousand DOGs using wide-area surveys or deep fields (e.g., Toba \& Nagao \citeyear{Toba16}; Yu et al. \citeyear{Yu24}, Noboriguchi et al. \citeyear{Nobo25}).
However, most of DOGs found in those earlier surveys have not been classified into subclasses (bump and PL DOGs) due to the lack of deep multi-band photometry, particularly in the NIR. 
So far, only a few hundred DOGs have been classified in previous studies, which is not sufficient to study the statistical properties for each subclass of DOGs (e.g., luminosity function).
To address this, we focus on the Ultraviolet Near-Infrared Optical Northern Survey (UNIONS; Gwyn et al. \citeyear{Gwyn25}) and $Euclid$ ($Euclid$ Collaboration \citeyear{Euclid24}).
UNIONS is an ongoing multi-band optical imaging survey aiming to obtain deep $ugriz$ images over $\sim$ 5,000 deg$^2$ of the northern sky, which mostly overlaps with the northern part of the $Euclid$ survey area.
$Euclid$ is a space telescope that is currently carrying out single-band optical and multi-band NIR imaging survey with high sensitivity. 
By combining UNIONS and $Euclid$ data, we will be able to construct a statistically large DOG sample with deep multi-band photometry in the near future.
This will allow us to study statistical properties for each subclass in detail.

As a preparatory step toward UNIONS-$Euclid$ studies on DOGs, in this paper, we select DOGs using preliminary UNIONS data combined with the UKIRT Infrared Deep Sky Survey (UKIDSS; Lawrence et al. \citeyear{Lawrence07}) instead of $Euclid$. 
This is because the $Euclid$ survey currently covers only a small fraction of its planned survey area.
In this paper, we present SED analyses of 376 DOGs and report the discovery of a new subclass of AGN-dominated DOGs.
This paper is organized as follows: Section \ref{sec:sample} presents the sample selection of DOGs.
We classify them into bump DOGs and PL DOGs, and describe their basic properties in Section \ref{sec:results}.
In Section \ref{sec:discussion}, we discuss the nature of the new subclass DOGs and compare them with other dusty AGNs. 
The cosmology adopted in this paper assumes a flat universe with $H_{0} = 70$ kms$^{-1}$ Mpc$^{-1}$, $\Omega _{M} = 0.3$, and $\Omega _{\Lambda} =0.7$. 
Unless otherwise noted, all magnitudes refer to the AB system.

\section{Data and Sample Selection} \label{sec:sample}

Using the photometric data from optical through NIR to MIR wavelengths, we identified 376 DOGs over $\sim$ 170 deg$^2$ following the manner adopted by Toba et al. (\citeyear{Toba15}; see also Noboriguchi et al. \citeyear{Nobo19}). 
Details of the photometric catalogs and the DOG selection process are provided as follows.

\subsection{UNIONS sample} \label{subsec:UNIONS}

UNIONS is a collaboration of wide field imaging surveys (will reach $\sim$ 5000 deg$^2$) of the northern hemisphere. UNIONS consists of Canada-France Imaging Survey (CFIS; Ibata et al. \citeyear{Ibata17}) conducted at the 3.6-meter CFHT on Maunakea, Panoramic Survey Telescope and Rapid Response System (Pan-STARRS; Chambers et al. \citeyear{Chambers16}), Wide Imaging with Subaru Hyper Suprime-Cam (HSC; Miyazaki et al. \citeyear{Miyazaki18}) of the $Euclid$ Sky (WISHES), and Waterloo-Hawaii Institute for Astronomy $g$-band Survey (WHIGS). 
CFHT/CFIS is obtaining deep $u$ and $r$ bands, Pan-STARRS is obtaining deep $i$-band and moderate-deep $z$-band imaging, and Subaru is obtaining deep $z$-band imaging through WISHES and $g$-band imaging through WHIGS. 
These independent efforts are directed, in part, to securing optical imaging to complement the $Euclid$ space mission, although UNIONS is a separate collaboration aimed at maximizing the science return of these large and deep surveys of the northern skies.
The 10$\sigma$ detection limits in a 2$^{\prime\prime}$ diameter aperture are 23.7, 24.5, 24.2, 23.8, and 23.3 magnitudes for the $u$, $g$, $r$, $i$, and $z$ bands, respectively (Gwyn et al. \citeyear{Gwyn25}).

For the photometry, UNIONS employs the ``GAaP'' pipeline, designed to provide consistent color measurements for $r$-band detected sources across different telescopes (Kuijken et al. \citeyear{Kuijken15}, \citeyear{Kuijken19}). 
The photometric aperture size is determined by the major and minor axes (\texttt{Agaper} and \texttt{Bgaper}, respectively) using the following equation:
\begin{equation}
\label{eq: photo}
\begin{split}
X{\rm gaper} = (X\_{\rm WORLD}^2 + {\rm MIN\_APER}^2)^{1/2} \\
\quad\text{for $X = A, B$}
\end{split}
\end{equation}
where $X\_{\rm WORLD}$ and ${\rm MIN\_APER}$ are the SExtractor size and shape parameters, respectively (Bertin \& Arnouts \citeyear{Bertin96}). The ${\rm MIN\_APER}$ parameter has two possible values: 0.7 and 1.0 arcseconds (see Kuijken et al. \citeyear{Kuijken15}, \citeyear{Kuijken19} for more details).
The photometric aperture size for all five bands is defined using Equation (\ref{eq: photo}) based on the CFIS $r$-band.

In this study, we used an internal data release of the UNIONS data (``UNIONS2000'' of UNIONS Grand Unified Catalog). 
For the magnitudes of each sources, we adopted  the ``optimal magnitude'', which choice the aperture size with the optimal ${\rm MIN\_APER}$ value (i.e., 0.7 or 1.0 arcsec; see Kuijken et al. \citeyear{Kuijken19} for more details). 
Additionally, we selected objects with ``\texttt{FLAG\_GAAP\_i} = 0'' (indicating successful GAaP photometry) and ``$i$-band magnitude $\neq 99$'' (detected in the $i$ band) and ``S/N ($i$ band) $\geq 3$'' to construct a reliable UNIONS sample.

\subsection{UKIDSS sample} \label{subsec:UKIDSS}

We obtained the NIR photometric data based on UKIDSS. 
Specifically, we used UKIDSS Large Area Survey (LAS) DR11plus.
The 5$\sigma$ detection limits (2$^{\prime\prime}$ aperture) are 20.2, 19.6, 18.8, and 18.2 Vega magnitudes in the $Y$, $J$, $H$, and $K$ bands, respectively\footnote{\url{http://wsa.roe.ac.uk/dr11plus_release.html}}.
The offset values between Vega and AB system are 0.634, 0.938, 1.379, and 1.900 in the $Y$, $J$, $H$, and $K$ bands, respectively, following Hewett et al. (\citeyear{Hewett06}).
To construct a reliable UKIDSS sample, we selected objects with \texttt{priOrSec} = (0 or \texttt{frameSetID}) (select unique sources), \texttt{kppErrBits} = 0 (no error bits in $K$ band), and S/N ($K$ band) $\geq$ 3.

\subsection{WISE sample} \label{subsec:WISE}

We obtained MIR data based on the Wide-field Infrared Survey Explorer (WISE; Wright et al. \citeyear{Wright10}). 
Specifically, we used AllWISE catalogs (Wright et al. \citeyear{Wright19}; Cutri et al. \citeyear{Cutri21}). 
The 5$\sigma$ detection limits of the AllWISE are 0.054, 0.071, 0.73, and 5 mJy at 3.4, 4.6, 12, and 22 $\mu$m, respectively\footnote{\url{https://wise2.ipac.caltech.edu/docs/release/allwise/expsup/sec2_3a.html}}.
%The offset values between Vega and AB system are 2.699, 3.339, 5.174, and 6.620 for 3.4, 4.6, 12, and 22 $\mu$m, respectively\footnote{\url{https://wise2.ipac.caltech.edu/docs/release/allsky/expsup/sec4_4h.html}}.
To construct a reliable WISE sample, we selected objects with \texttt{w4\_sat} = 0 (no saturation pixel at 22 $\mu$m), \texttt{w4\_cc\_map} = 0 (no image artifacts at 22 $\mu$m), \texttt{ext\_flg} = 0 (not extended), and S/N (22 $\mu$m) $\geq$ 3.

\subsection{Selection of DOGs} \label{subsec:DOGs}

First, we cross-matched the reliable UNIONS and UKIDSS samples by adopting a search radius of 1$^{\prime\prime}$. 
As a result, we identified 1,460,178 sources in the overlap region between the UNIONS and UKIDSS LAS survey area ($\sim$ 170 deg$^2$). 
Before performing UNIONS-WISE cross-matching, we excluded sources unlikely to be DOGs in order to prevent mismatches due to the significant difference in the angular resolution between UNIONS and WISE. 
Specifically, we selected red objects that meet the criterion of $i - K \geq 1.2$, leaving 874,151 sources.

Second, we cross-matched the above UNIONS-UKIDSS matched sample with the reliable WISE sample by adopting a search radius of 3$^{\prime\prime}$. 
As a result, 8,438 sources remained. 

Finally, we adopted the definition of the DOG as follows:
\begin{equation}
(i - [22])_{\rm AB} \geq 7 
\end{equation}
\noindent 
where $i$ and [22] represent the magnitudes in the UNIONS $i$-band and WISE 22 $\mu$m band, respectively (Toba et al. \citeyear{Toba15}; see also Dey et al. \citeyear{Dey08})
%% ==========
%%  footnote 
%% ==========
\footnote{
Note that the selection criterion defined by Toba et al. (\citeyear{Toba15}) does not  take a redshift dependence into account. 
Actually, some authors reported local DOGs by considering rest-frame DOG color (e.g., Hwang \& Geller \citeyear{Hwang13}).
On the other hand, our criterion is optimized for DOG search with HSC and WISE, where traditional DOGs in a wide redshift range down to $z \sim 1$ also satisfy the criterion (e.g., Toba et al. \citeyear{Toba15}; Noboriguchi et al. \citeyear{Nobo19}).}. 
%% =====
%%  end
%% =====
Using this criterion, we successfully identified 382 DOGs over $\sim$ 170 deg$^2$.  
However, we excluded 6 objects that were undetected at both 4.6 and 12 $\mu$m, as these could not be classified into any subclasses (see Section \ref{subsec:subclass}).  
Therefore, we focus on the remaining 376 DOGs in the following analyses. 
Note that we used only the $u$, $g$, $r$, and $i$-band  UNIONS photometry to study their SEDs, because the $z$-band photometry of UNIONS2000 is not currently available in the overlap region of UNIONS and UKIDSS.

\section{Results} \label{sec:results}

\subsection{22 $\mu$m flux} \label{subsec:22 mum}

%% ========
%%  figure 
%% ========
\begin{figure}[t]
\includegraphics[width=8.5cm]{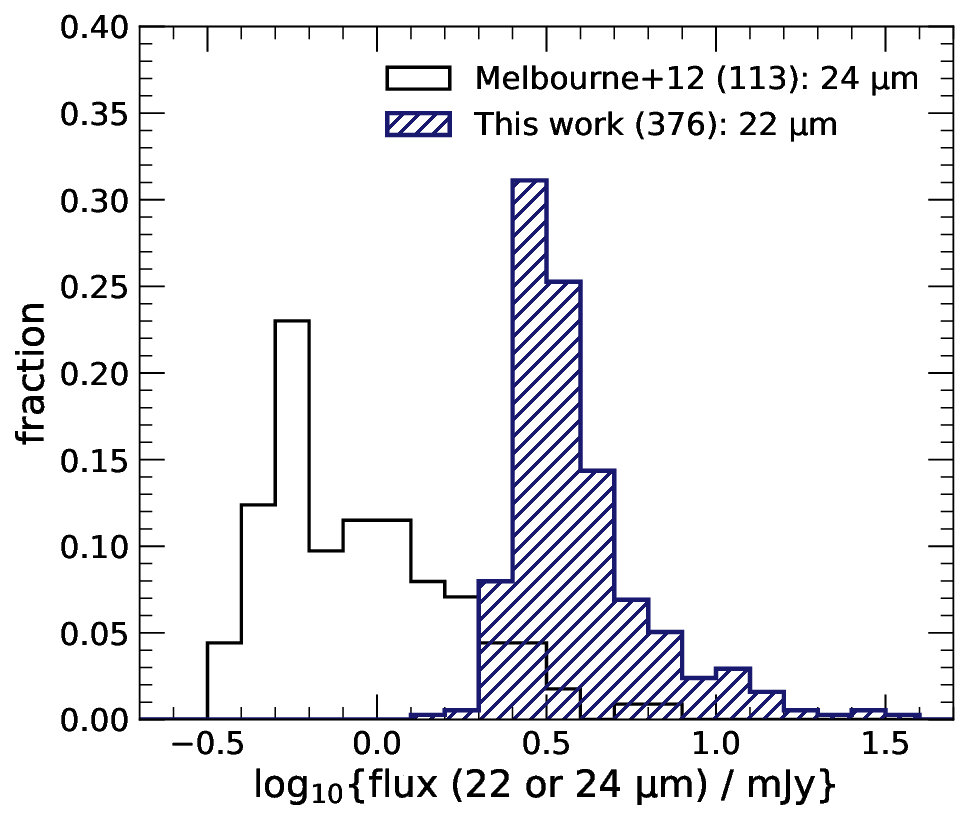}
\caption{
Comparison of the 22 $\mu$m flux distribution of our DOG sample with the 24 $\mu$m flux distribution of DOGs from Melbourne et al. (\citeyear{Melbourne12}). 
The numbers in parentheses indicate the number of objects in each sample.
\label{fig: IR}}
\end{figure}
%% =====
%%  end
%% =====

In some past studies, DOGs were identified in deep fields with a high sensitivity but consequently with a narrow survey area, as those studies focused on faint DOGs. 
In contrast, we combined wide-field optical and NIR survey data with the AllWISE catalog covering a largest survey area but with a shallower sensitivity. 
As a result, our DOG sample has higher MIR fluxes compared to DOGs constructed in deep fields. 
Such DOGs are referred to as IR-bright DOGs (Toba et al. \citeyear{Toba15}; Toba \& Nagao \citeyear{Toba16}; Toba et al. \citeyear{Toba17c}, \citeyear{Toba17a}; Noboriguchi et al. \citeyear{Nobo19}) and typically have lower redshifts ($z \sim 1$; Toba et al. \citeyear{Toba17c}; Noboriguchi et al. \citeyear{Nobo19}; but see also Toba \& Nagao \citeyear{Toba16}; Toba et al. \citeyear{Toba15}) than relatively faint DOGs in deep fields ($z \sim 2$; Dey et al. \citeyear{Dey08}; Riguccini et al. \citeyear{Riguccini11}; Yu et al. \citeyear{Yu24}).

Figure \ref{fig: IR} compares the 22 $\mu$m flux of our DOG sample with the 24 $\mu$m flux of DOGs in the NOAO Bo\"{o}tes field, as reported by Melbourne et al. (\citeyear{Melbourne12}). 
Our DOG sample exhibits fluxes (median flux: 3.37 [mJy]) that are $\sim$ 0.6 dex brighter than those in the Melbourne et al. (\citeyear{Melbourne12}) sample (median flux: 0.80 [mJy]). 

Toba et al. (\citeyear{Toba15}) reported a positive correlation between 22 $\mu$m flux and the AGN activity.
Therefore, this systematically higher MIR flux suggests that our DOGs harbor more active AGN systems than past samples obtained in deep fields. 
This is important for understanding the co-evolution between SMBHs and their host galaxies.

\subsection{Subclasses and spectral energy distributions} \label{subsec:subclass}

%% ========
%%  figure 
%% ========
\begin{figure}[t]
\includegraphics[width=8.5cm]{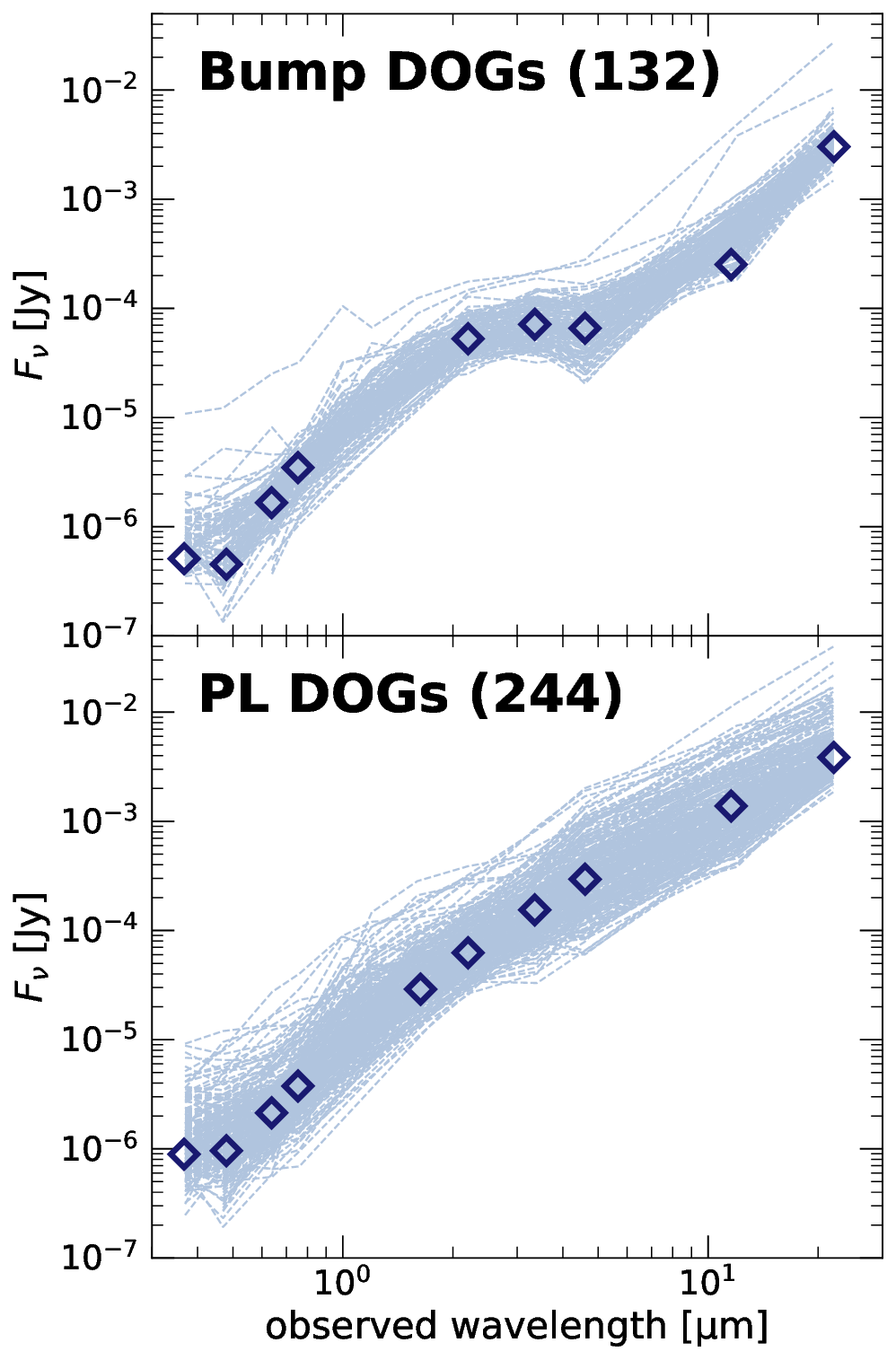}
\caption{
SEDs of bump DOGs (top panel) and PL DOGs (bottom panel). The diamonds and dashed lines represent the median and individual SEDs, respectively. 
Note that the median flux density is not shown for bands in which more than half of the objects are undetected.
The numbers in parentheses indicate the number of objects in each subclass.
\label{fig: suclass}}
\end{figure}
%% =====
%%  end
%% =====

Based on their SEDs, we classified our DOGs into two subclasses following the same manner as Toba et al. (\citeyear{Toba15}; see also Noboriguchi et al. \citeyear{Nobo19}). 
First, we fitted the MIR slope ($\alpha _{\rm MIR}$) of their SEDs with a power-law model ($F _{\nu} \propto \lambda ^{\alpha}$) from 4.6 to 22 $\mu$m and estimated NIR $K$-band flux based on this power-law fit ($f^{\rm fit}_{K}$).
Next, we compared the estimated $K$-band flux ($f^{\rm fit}_{K}$) with the observed $K$-band flux ($f^{\rm obs}_{K}$).
DOGs were classified as bump DOGs when they satisfied the following condition:
\begin{equation}
f^{\rm obs}_{K}/f^{\rm fit}_{K} \geq 3.
\end{equation}
This flux ratio is a reliable indicator to identify the bump feature (see Toba et al. \citeyear{Toba15}).

As a result, $\sim$ 35\% (132/376) and $\sim$ 65\% (244/376) of our DOGs were classified as bump DOGs and PL DOGs, respectively. 
This fraction is roughly consistent with previous works on IR-bright DOGs (e.g., 17/57 $\sim$ 35\% and 31/57 $\sim$ 65\% objects were classified as bump and PL DOGs in Toba et al. \citeyear{Toba15}). 
Figure \ref{fig: suclass} shows the obtained SEDs of the bump DOGs and PL DOGs, respectively.
The peak of bump feature in the SED of bump DOGs is shown around 3.4 $\mu$m, suggesting that their redshifts are $\sim$ 1. 
This is consistent with the typical redshift for IR-bright DOGs (see Section \ref{subsec:22 mum}).

\subsection{Comparison of slopes} \label{subsec:slope}

%% ========
%%  figure 
%% ========
\begin{figure}[t]
\includegraphics[width=8.5cm]{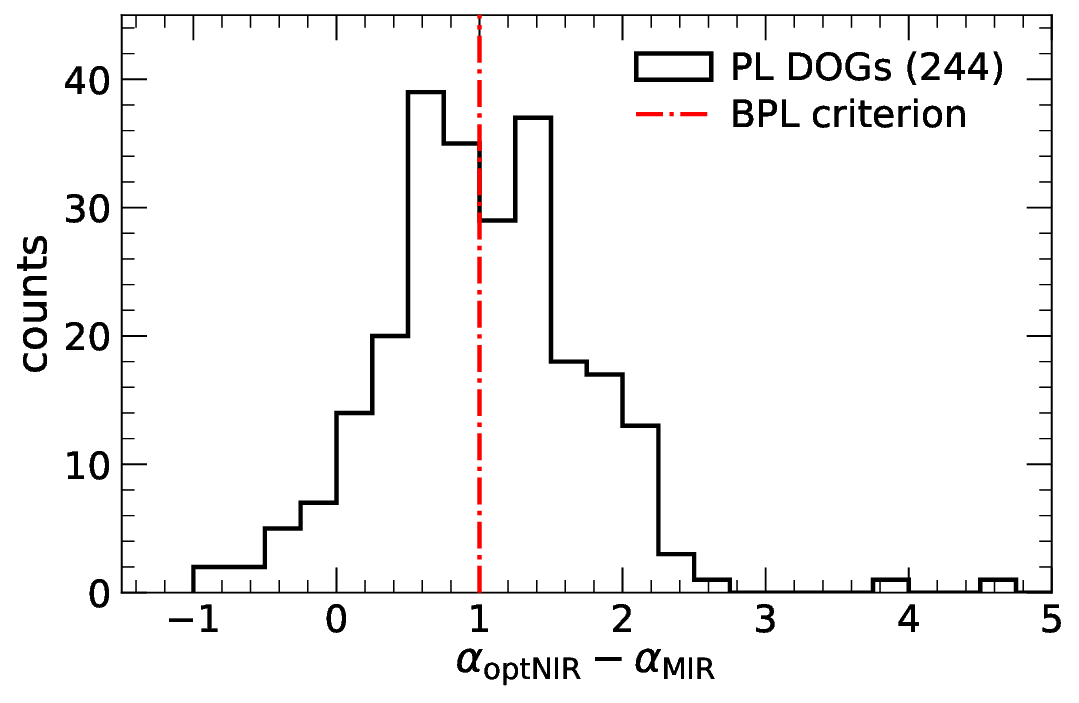}
\caption{
The histogram of the difference between the optical-NIR and MIR slopes ($\alpha _{\rm optNIR} - \alpha _{\rm MIR}$) of PL DOGs.
The red dash-dotted line represents the criterion to separate normal-PL (NPL) and broken-PL (BPL) DOGs. 
The number in parentheses indicates the total count of PL DOGs.
\label{fig: slope_dif}}
\end{figure}
%% =====
%%  end
%% =====

During the SED analysis, we found that a number of PL DOGs show ``broken'' power-law SEDs. 
The power-law shapes in the SEDs of these ``broken power-law DOGs'' (hereafter BPL DOGs) are broken at $\lambda _{\rm obs} \sim 4$ $\mu$m, unlike those of normal power-law DOGs (hereafter NPL DOGs). 
A break seen in the SED of BPL DOGs indicates the presence of a significant difference in the slopes of the SED between shorter and longer wavelengths.
To quantitatively evaluate the difference in slopes between shorter and longer wavelengths, we additionally fitted their optical-NIR slopes ($\alpha _{\rm optNIR}$) by a power law from $g$-band to $K$-band, and compared it with their $\alpha _{\rm MIR}$ (see Section \ref{subsec:subclass}).

%% ========
%%  figure 
%% ========
\begin{figure*}[t]
\includegraphics[width=17.7cm]{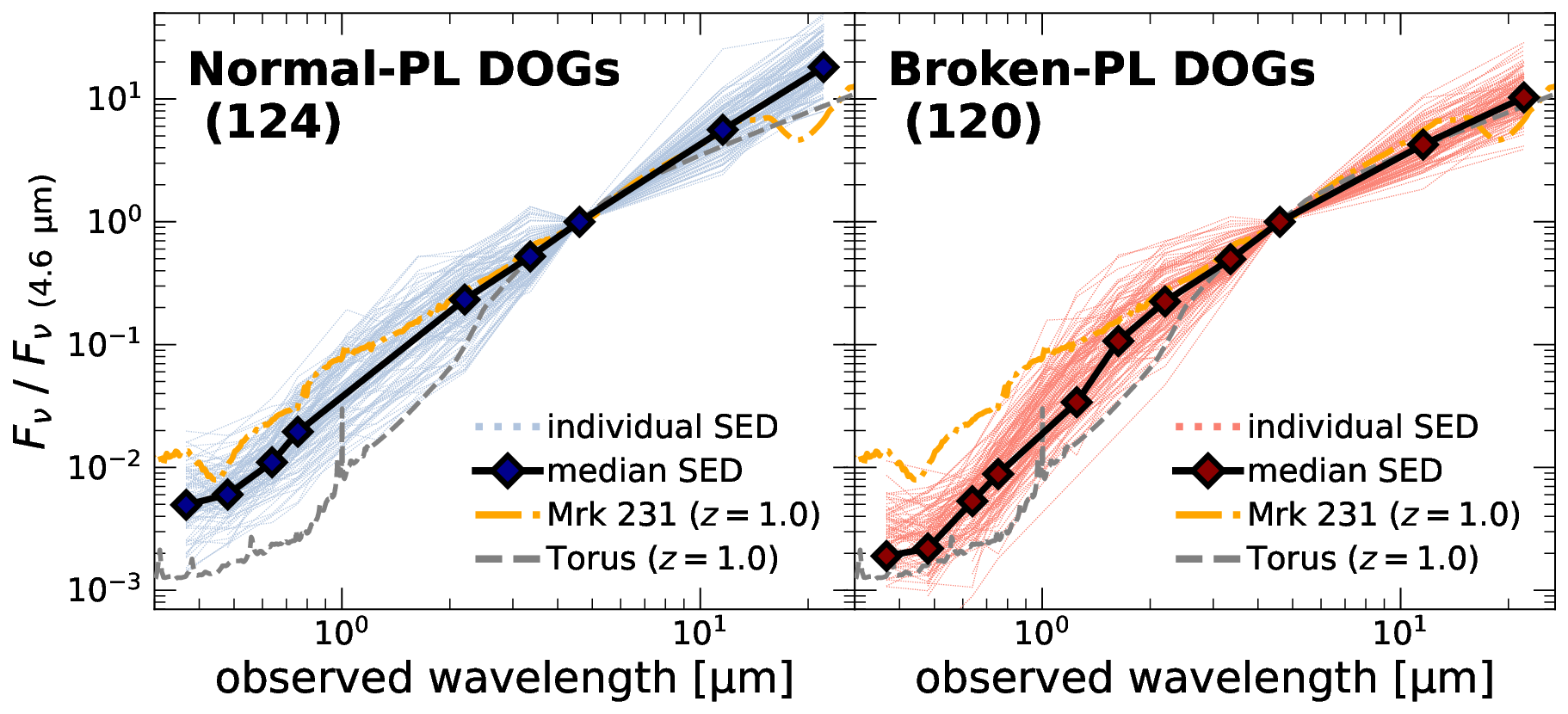}
%\plotone{BPL_NPL.eps}
\caption{
SEDs of normal-PL DOGs (left panel) and broken-PL DOGs (right panel). 
The diamonds and dotted lines represent the median and individual SEDs, respectively. 
Note that the median flux density is not shown for bands in which more than half of the objects are undetected. 
The orange dash-dotted lines and gray dashed lines show the SED template of Mrk 231 and torus model from Polletta et al. (\citeyear{Polletta06}, \citeyear{Polletta07}), respectively, assuming $z=1$. 
SEDs are normalized at $\lambda _{\rm obs} = 4.6$ $\mu$m (around the break of the broken-power-law SED shape; see Section \ref{subsec:slope}).
The numbers in parentheses indicate the number of objects in each subclass.
\label{fig: NPL-BPL}}
\end{figure*}
%% =====
%%  end
%% =====

Figure \ref{fig: slope_dif} shows the difference between the optical-NIR and MIR slopes ($\alpha _{\rm optNIR} - \alpha _{\rm MIR}$) of PL DOGs.
The mean and standard deviation of $\alpha _{\rm optNIR} - \alpha _{\rm MIR}$ for 244 PL DOGs are 1.0 and 0.7, respectively.
The distribution of $\alpha _{\rm optNIR} - \alpha _{\rm MIR}$ appears to have a bimodal structure, suggesting the possible presence of two populations corresponding to NPL and BPL DOGs.
Based on this, we define BPL DOGs as PL DOGs with $\alpha _{\rm optNIR} - \alpha _{\rm MIR} \geq 1.0$, which corresponds to the local minimum between the two potential peaks. 
As a result, roughly half (120/244) of the PL DOGs are classified as BPL DOGs. 
This large fraction of BPL DOGs among PL DOGs possibly reflects a selection bias, which will be discussed in Section \ref{subsec:k-band}. 
We show the individual and median SEDs of NPL and BPL DOGs in Figure \ref{fig: NPL-BPL}. 
This figure shows that $\alpha_{\rm MIR}$ is not significantly different between NPL and BPL DOGs, while BPL DOGs show steeper $\alpha_{\rm optNIR}$ than NPL DOGs.

%% ========
%%  figure 
%% ========
\begin{figure}[t]
\includegraphics[width=8.5cm]{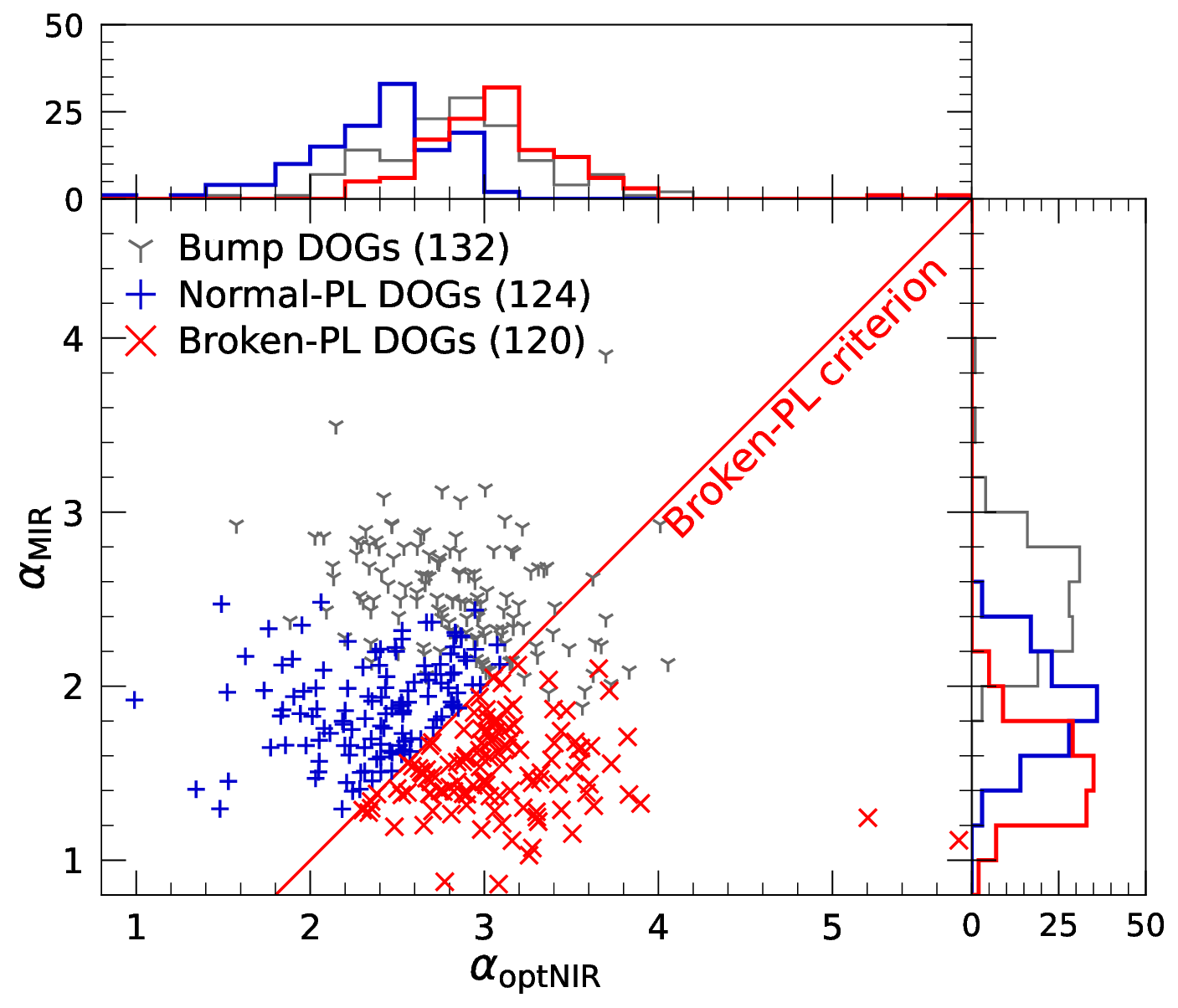}
\caption{
The comparison of optical-NIR slopes ($\alpha _{\rm optNIR}$) and MIR slopes ($\alpha _{\rm MIR}$) for each of the subclasses (black Ys: bump DOGs, blue pluses: normal-PL DOGs, red crosses: broken-PL DOGs). 
The red solid line represents the criterion of broken-PL DOGs. 
The numbers in parentheses indicate the number of objects in each subclass. 
The black, blue, and red histograms show the distribution of $\alpha _{\rm optNIR}$ (top panel) and $\alpha _{\rm MIR}$ (right panel) for bump, normal-PL, broken-PL DOGs, respectively. 
\label{fig: slope}}
\end{figure}
%% =====
%%  end
%% =====

Figure \ref{fig: slope} shows the comparison of $\alpha _{\rm optNIR}$ and $\alpha _{\rm MIR}$ of three subclasses of DOGs.
First, bump and PL DOGs (including both broken and normal) are separated by $\alpha _{\rm MIR}$, consistent with the results of Noboriguchi et al. (\citeyear{Nobo19}).
They reported that bump DOGs have redder colors in the MIR compared to PL DOGs (see Section 2.1.3 and Figure 2 of Noboriguchi et al. \citeyear{Nobo19} for more details), which is presumably attributed to the hot dust around the AGN in PL DOGs. 
Second, PL DOGs have a large scatter in their $\alpha _{\rm optNIR}$ with a standard deviation of 0.56, compared to their $\alpha _{\rm MIR}$ whose standard deviation is 0.32.
These points will be discussed in the following section.

Finally, we performed Kolmogorov–Smirnov test for $\alpha _{\rm optNIR}$ and $\alpha _{\rm MIR}$ between each pair of subclasses (Table \ref{tab: KS}). 
The $p$-values are significantly small for all cases, indicating that the distributions of the subclasses in the $\alpha _{\rm optNIR}$-$\alpha _{\rm MIR}$ plane are statistically different with $>3 \sigma$ significance. 
%% =======
%%  table 
%% =======
\begin{table}[t]
\caption{The $p$-value of Kolmogorov–Smirnov test for $\alpha _{\rm optNIR}$ and $\alpha _{\rm MIR}$ between each pair of subclasses.
\label{tab: KS}}
\centering
\begin{tabular}{ccc}
\hline
Pair of subclasses & $\alpha _{\rm optNIR}$ & $\alpha _{\rm MIR}$ \\
\hline
\hline
bump vs normal-PL & $9.1 \times 10^{-13}$ & $1.4 \times 10^{-30}$ \\
bump vs broken-PL & $2.4 \times 10^{-4}$ & $8.0 \times 10^{-64}$ \\
normal-PL vs broken-PL & $9.3 \times 10^{-26}$ & $3.5 \times 10^{-16}$ \\
\hline
\end{tabular}
\end{table}
%% =====
%%  end
%% =====

\section{Discussion} \label{sec:discussion}

\subsection{BPL DOGs as heavily obscured AGNs} \label{subsec:heavy}
In Section \ref{subsec:slope}, we showed that PL DOGs have a large scatter in their $\alpha _{\rm optNIR}$, compared to their $\alpha _{\rm MIR}$. 
This result is consistent with the results of Banerji et al. (\citeyear{Banerji13}), who performed a similar power-law fit (at rest-frame 0.3--1 $\mu$m and 1--3 $\mu$m) to heavily reddened type-1 quasars.
They showed that heavily reddened type-1 quasars show a large scatter in the spectral slope at shorter wavelength. 
However, in the longer wavelength, their spectral slope does not show such a large scatter.
They attribute the large scatter in the spectral slope at the shorter wavelength to the diversity in the amount of dust extinction (see Section 4 of Banerji et al. \citeyear{Banerji13} for more details). 
Given the similarity in the scatter of the SEDs between the sample of Banerji et al. (\citeyear{Banerji13}) and ours, the large scatter in $\alpha _{\rm optNIR}$ of our PL DOGs is also consistent with the large scatter of the amount of dust extinction.
Based on the above considerations, BPL DOGs are plausibly more heavily obscured AGNs than NPL DOGs.
However, a major question remains; what causes this difference in extinction between the two populations of PL DOGs?
To address this question, we propose two possible scenarios as follows.

\subsubsection{Viewing angle scenario} \label{subsubsec:viewing}

One possible scenario is that the difference between NPL and BPL DOGs depends on the viewing angle like the type1/2 dichotomy of AGNs.
That is, NPL DOGs are seen from a face-on angle, while BPL DOGs are seen from an edge-on angle.
In other words, BPL DOGs are affected by both host extinction and torus extinction, while NPL DOGs are affected only by host extinction.
In this scenario, bump DOGs evolve into either NPL or BPL DOGs depending on the viewing angle.

Toba et al. (\citeyear{Toba17c}) studied the optical spectra of IR-bright DOGs using the Sloan Digital Sky Survey (SDSS; York et al. \citeyear{York20}) and WISE data. 
They had no NIR photometry and thus could not classify their DOGs sample into bump and PL DOGs.
However, most of their sample are probably PL DOGs, because of the high 22 $\mu$m flux of their IR-bright DOGs sample (see also Noboriguchi et al. \citeyear{Nobo19}). 
They spectroscopically identified both type-1 and type-2 AGNs in their sample of IR-bright DOGs, but they could not study possible breaks in the SEDs of their sample due to the lack of NIR photometric data. 
The type-2 AGNs in their sample of PL DOGs may correspond to BPL DOGs, but the NIR photometry is needed to confirm this (see also Brand et al. \citeyear{Brand07}; Ross et al. \citeyear{Ross15} for spectroscopic properties of related dusty AGNs with the same color selection as DOGs).

Returning to Figure \ref{fig: NPL-BPL}, we compared the SED shapes of NPL and BPL DOGs with the SED templates of Mrk 231 and torus model (Polletta et al. \citeyear{Polletta06}, \citeyear{Polletta07}) as heavily obscured type-1 and type-2 quasars, respectively.
The redshift of these SED models is assumed to be $z=1.0$, which is the typical for IR-bright DOGs (see Section \ref{subsec:22 mum}). 
The SED model of the Mrk 231 looks non-broken power-law, while the torus model have a break and the same $\alpha _{\rm optNIR}$ in the SED. 
These features are similar to NPL and BPL DOGs, respectively, which supports this scenario.

\subsubsection{Transition phase scenario} \label{subsubsec:transition}

Another possible scenario is that BPL DOGs represent an earlier evolutionary stage than NPL DOGs in the PL DOG phase.
In Section \ref{subsec:heavy}, we discussed the similarity of heavily reddened type-1 quasars in Banerji et al. (\citeyear{Banerji13}) and our sample of PL DOGs. 
Here we compare the median of slopes between these sample to quantitatively evaluate the similarity. 
The median of slopes (in the optical-NIR and MIR) are similar between their heavily reddened type-1 quasar sample (2.47 and 1.57)
%% ==========
%%  footnote 
%% ==========
\footnote{In Banerji et al. (\citeyear{Banerji13}), they fitted the slopes by power laws of the form $\nu F _{\nu} \propto \nu ^{\beta}$. 
Therefore, we converted their $\beta$ to our $\alpha$ using $\alpha = - \beta +1$.}
%% =====
%%  end
%% =====
 and our PL DOG sample (2.72 and 1.68).
This similarity suggests that the difference in reddening of PL DOGs can be explained even without torus obscuration. 
Therefore BPL DOGs may be chracterized by more heavily obscuration by dust at the spatial scale of host galaxies, than NPL DOGs. 
Considering that the powerful radiation and/or outflow from the AGN blows out such large-scale dust, BPL DOGs likely represent an earlier evolutionary stage than NPL DOGs in the PL DOG phase. 
Additionally, when combined with the gas-rich major merger scenario, BPL DOGs may represent the transitional phase in the evolution from bump DOGs to NPL DOGs.

\subsection{Correlation between $\alpha _{\rm optNIR} - \alpha _{\rm MIR}$ and $K$-band flux} \label{subsec:k-band}

In Section \ref{subsec:slope}, we found that our sample has the large fraction of BPL DOGs among PL DOGs (120/244 $\sim$ 49\%). 
However, it remains uncertain whether this large fraction is also present in other DOG samples made by earlier observational studies or not.
To check this, we searched BPL DOGs in the sample of Noboriguchi et al. (\citeyear{Nobo19}). 
They studied IR-bright DOGs using the data from Subaru HSC, the VISTA Kilo-degree Infrared Galaxy survey (VIKING; Edge et al. \citeyear{Edge13}), and WISE, providing deeper photometry in optical and NIR than ours. 
We fitted their slopes with the same manner as ours to infer $\alpha _{\rm optNIR}$ and $\alpha _{\rm MIR}$
%% ==========
%%  footnote 
%% ==========
\footnote{There is a difference in the classification of bump and PL DOGs between Noboriguchi et al. (\citeyear{Nobo19}) and our sample. 
Noboriguchi et al. (\citeyear{Nobo19}) classified objects detected in only two among 3 bands (4.6, 12, and 22 $\mu$m) as ``unclassified'' DOGs. 
In contrast, we classified such objects into bump or PL DOGs following the method used by Toba et al. (\citeyear{Toba15}). 
For a fair comparison, we re-classified those uncassified DOGs in the Noboriguchi et al. (\citeyear{Nobo19}) sample using the same criteria.}. 
%% =====
%%  end
%% =====
We found that 24\% (74/300) of the PL DOGs in the sample of Noboriguchi et al. (\citeyear{Nobo19}) are classified as BPL DOGs, which is approximately half the fraction observed in our sample. 

%% ========
%%  figure 
%% ========
\begin{figure}[t]
\includegraphics[width=8.5cm]{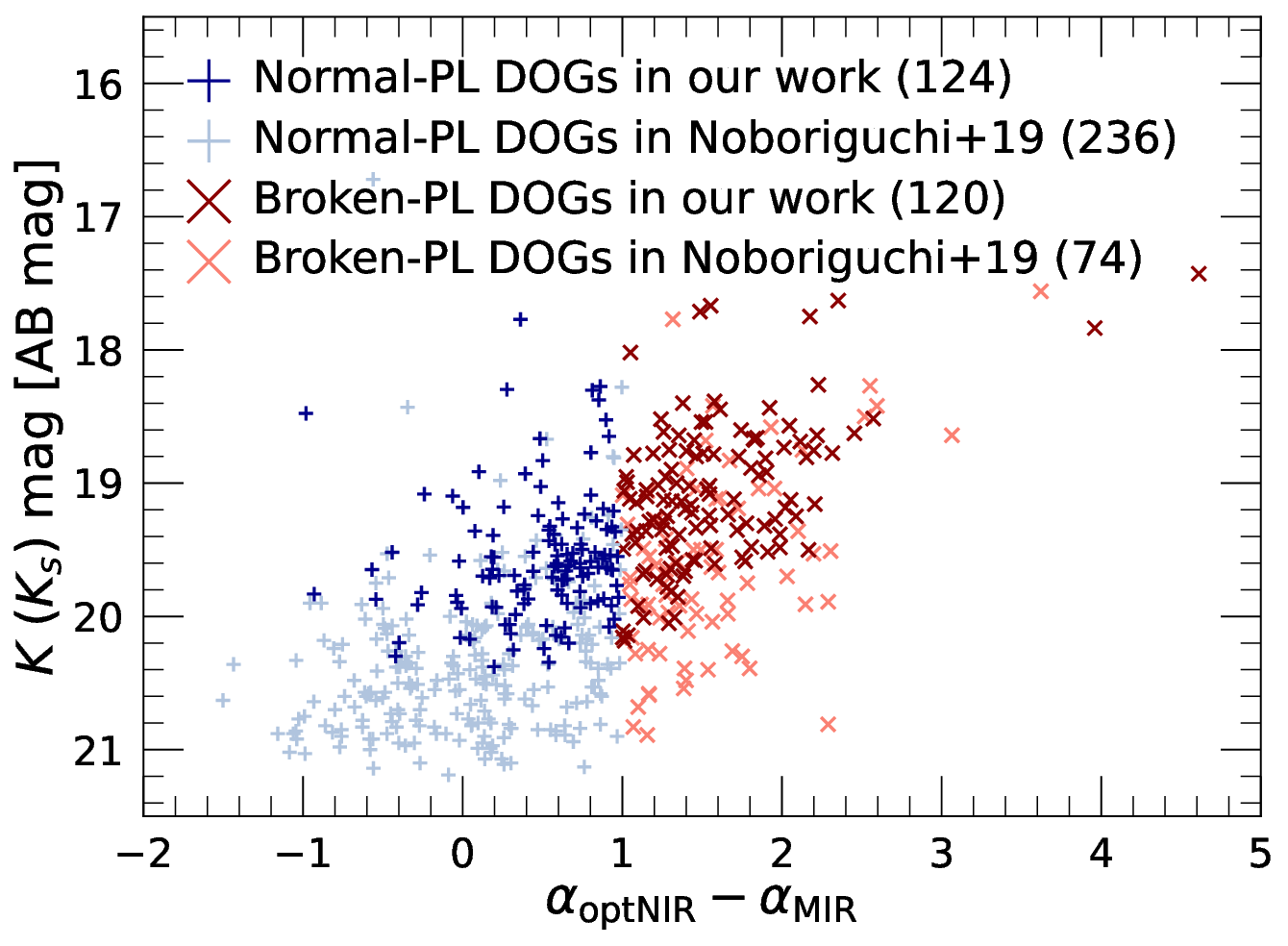}
\caption{
$K$-band magnitude of our sample and $K _{\rm s}$-band magnitude of Noboriguchi et al. (\citeyear{Nobo19}) sample as a function of the difference between the optical-NIR slope and MIR slope ($\alpha _{\rm optNIR} - \alpha _{\rm MIR}$). 
Two samples are shown: our sample and the sample from Noboriguchi et al. (\citeyear{Nobo19}), denoted by darker and lighter colors, respectively. 
Blue pluses and red crosses indicate normal-PL DOGs and broken-PL DOGs, respectively. 
The numbers in parentheses indicate the number of objects in each subclass in each sample.
\label{fig: bias}}
\end{figure}
%% =====
%%  end
%% =====

This lower fraction may explain why BPL DOGs have received little attention in previous studies, if such a low fraction in the sample of Noboriguchi et al. (\citeyear{Nobo19}) is a general property of BPL DOGs. 
But what causes this difference in the BPL fraction? 
One possible idea is that the difference in detection limits between the sample of Noboriguchi et al. (\citeyear{Nobo19}) and ours may be responsible for this discrepancy. 
To investigate this, we examined the relationship between the $K$-band (or $K_{\rm s}$-band for the sample of Noboriguchi et al. \citeyear{Nobo19}) magnitude and $\alpha_{\rm optNIR} - \alpha_{\rm MIR}$. 
We focused on the $K$-band magnitude because the difference in the NIR detection limit between the two samples is more significant than that in the optical. 
As a result, we found a correlation between the $K$-band (or $K_{\rm s}$-band) magnitude and $\alpha_{\rm optNIR} - \alpha_{\rm MIR}$ (see Figure \ref{fig: bias}).
To quantitatively assess the statistical significance of this correlation, we performed Spearman's rank correlation test using both the PL DOG sample from Noboriguchi et al. (\citeyear{Nobo19}) and our own sample. 
As a result, we obtained the correlation coefficient of $-0.63$ and the $p$-value of $2.1 \times 10^{-61}$, indicating the high statistical significance. 
This result suggests that the shallower detection limit of UKIDSS compared to that of VIKING preferentially selects PL DOGs that have large $\alpha _{\rm optNIR} - \alpha _{\rm MIR}$ with $K$-band ($K _{\rm s}$-band) flux. 
In other words, the larger fraction of BPL DOGs in our sample can be attributed to the shallower detection limit of UKIDSS compared to VIKING.

%% ========
%%  figure 
%% ========
\begin{figure}[t]
\includegraphics[width=8.5cm]{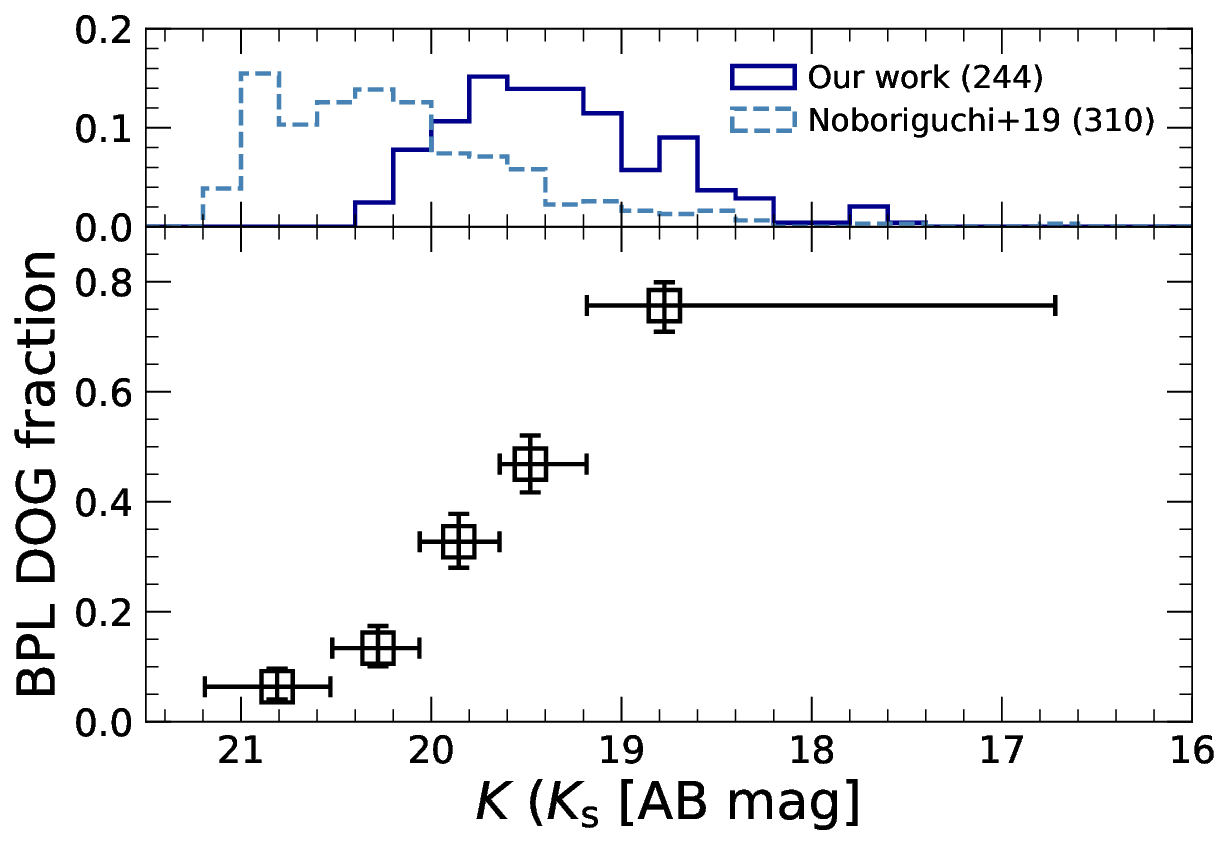}
\caption{
Fraction of broken-PL DOGs among PL DOGs (BPL DOG fraction) as a function of $K$-band ($K_{\rm s}$-band) magnitude (bottom panel), which is calculated based on the PL DOG samples in both Noboriguchi et al. (\citeyear{Nobo19}) and our work. 
The Poisson error in the fraction was estimated using the binomial statistics (see Gehrels \citeyear{Gehrels86}). 
Upper panel shows the distribution of $K$-band and $K_{\rm s}$-band magnitude for PL DOGs of our work (dark blue) and Noboriguchi et al. (\citeyear{Nobo19}) (light blue), respectively.
The numbers in parentheses indicate the counts of PL DOGs in each sample.
\label{fig: frac}}
\end{figure}
%% =====
%%  end
%% =====

%% =======
%%  table 
%% =======
\begin{table}[t]
\caption{Fraction of broken-PL DOGs among PL DOGs (BPL DOG fraction) as a function of $K$-band ($K_{\rm s}$-band) magnitude. 
\label{tab: frac}}
\centering
\begin{tabular}{cccc} % four columns, alignment for each
\hline
$K$ or $K_{\rm s}$ & $K$ or $K_{\rm s}$ & BPL DOG & Number of \\
range & median & fraction & PL DOGs \\
\hline
\hline
16.7--19.2 & 18.8 & $0.76_{-0.05}^{+0.04}$ & 111 \\
19.2--19.6 & 19.5 & $0.47_{-0.05}^{+0.05}$ & 111 \\
19.6--20.1 & 19.9 & $0.33_{-0.05}^{+0.05}$ & 110 \\
20.1--20.5 & 20.3 & $0.13_{-0.03}^{+0.04}$ & 112 \\
20.5--21.2 & 20.8 & $0.06_{-0.02}^{+0.03}$ & 110 \\
\hline
\end{tabular}
\end{table}
%% =====
%%  end
%% =====

To visualize this explanation, we calculated the fraction of BPL DOGs among PL DOGs in bins of $K$-band ($K_{\rm s}$-band) magnitude using the PL DOG samples in both Noboriguchi et al. \citeyear{Nobo19} and our work (see Table \ref{tab: frac}). 
The range of $K$-band ($K_{\rm s}$-band) magnitude for each bin was set so that the number of objects in each bin is as equal as possible. 
Figure \ref{fig: frac} shows the fraction of BPL DOGs among PL DOGs as a function of $K$-band ($K_{\rm s}$-band) magnitude. 
We confirmed that the fraction of BPL DOGs increases as the $K$-band ($K_{\rm s}$-band) magnitude decreases (i.e., as flux increases).

Assuming a typical redshift of IR-bright DOGs (i.e., $z \sim 1$), the $K$-band ($K _{\rm s}$-band) flux is affected by the rest-frame 1.6 $\mu$m bump, which traces SF activity. 
Thus, the higher $K$-band ($K _{\rm s}$-band) flux of BPL DOGs possibly indicates their higher SF activity compared to NPL DOGs.
This supports the transition phase scenario (Section \ref{subsubsec:transition}), assuming a gradual decrease of the SF activity from bump DOGs to NPL DOGs.

\subsection{Relationship with other dusty AGNs} \label{subsec:dusty}

\begin{figure}[t]
\includegraphics[width=8.5cm]{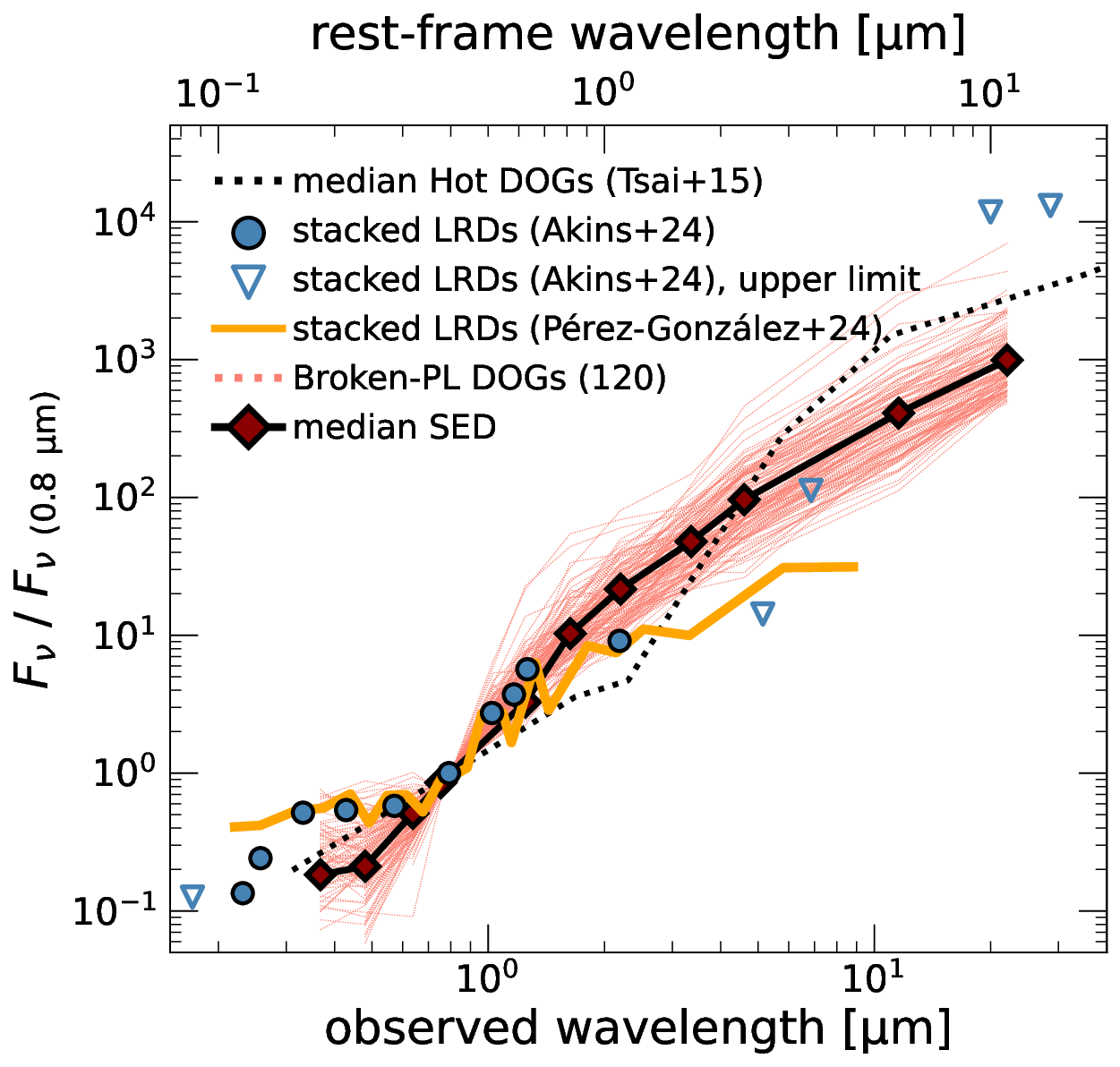}
\caption{
Comparison of the SEDs of broken-PL DOGs, the median SED of Hot DOGs (Tsai et al. \citeyear{Tsai15}), and the stacked SEDs of LRDs (P\'{e}rez-Gonz\'{a}lez et al. \citeyear{Perez24}; Akins et al. \citeyear{Akins24}; see also Li et al. \citeyear{Li25}).
The SEDs of Hot DOGs and LRDs are shifted to $z=1$, and normalized at $\lambda _{\rm obs} \sim 0.8$ $\mu$m. 
The texts and number in parentheses indicate the reference papers and the number of objects in our broken-PL DOG sample.
\label{fig: LRD}}
\end{figure}

Wu et al. (\citeyear{Wu12}) and Tsai et al. (\citeyear{Tsai15}) compared the SEDs of Hot DOGs with those of torus model (Polletta et al. \citeyear{Polletta06}, \citeyear{Polletta07}; see Section \ref{subsubsec:transition}) and found that the slopes of Hot DOGs are steeper than those of the torus SED model in the rest-frame optical wavelength, suggesting that Hot DOGs are more heavily obscured than the torus model. 
In Section \ref{subsubsec:viewing}, we showed that BPL DOGs have a similar $\alpha _{\rm optNIR}$ to the torus SED model, suggesting that Hot DOGs are the most heavily obscured AGNs among various DOG populations.
In Figure \ref{fig: LRD}, we compare the SEDs of our BPL DOGs with the median SED of Hot DOGs presented by Tsai et al. (\citeyear{Tsai15}). 
For a fair comparison, we shifted the median SED of Hot DOGs from their typical redshift of $z=3.0$ (Tsai et al. \citeyear{Tsai15}) to $z=1.0$, which is more representative of IR-bright DOGs.
Figure \ref{fig: LRD} shows a steeper slope of Hot DOG SED around rest-frame 1--5 $\mu$m range than those of BPL DOG SED, suggesting their heavier obscuration. 
The flatter SEDs of Hot DOGs in both shorter and longer wavelengths than those of BPL DOGs are due to their stellar emission and the excess of hot dust emission, respectively (e.g., Assef et al. \citeyear{Assef15}; Tsai et al. \citeyear{Tsai15}).

%% ========
%%  figure 
%% ========
%% =====
%%  end
%% =====

Next, we compared the SED of BPL DOGs with that of LRDs, which are candidates of high-$z$ ($z>4$) dust-reddened ($A_V \sim 3$ mag) type-1 AGNs. 
P\'{e}rez-Gonz\'{a}lez et al. (\citeyear{Perez24}) reported that LRDs have very red rest-frame optical slopes from $\lambda _{\rm rest} \sim$ 0.4 to 1 $\mu$m, which are similar to those of torus models (Polletta et al. \citeyear{Polletta06}, \citeyear{Polletta07}) as well as BPL DOGs. 
For the comparison, we shifted the stacked SEDs of LRDs to $z=1.0$, which is typical for IR-bright DOGs (see Section \ref{subsec:22 mum}). 
Figure \ref{fig: LRD} shows the comparison between the median and individual SEDs of BPL DOGs and the stacked SEDs of LRDs (P\'{e}rez-Gonz\'{a}lez et al. \citeyear{Perez24}; Akins et al. \citeyear{Akins24}; see also Li et al. \citeyear{Li25}). 
In P\'{e}rez-Gonz\'{a}lez et al. (\citeyear{Perez24}), they normalized the SEDs of LRDs at rest-frame 0.4 $\mu$m for comparison. 
Therefore we normalized SEDs at $\lambda _{\rm obs} \sim 0.8$ $\mu$m, assuming $z = 1.0$. 
The shapes of their SEDs are similar from $\lambda _{\rm obs} \sim$ 0.7 $\mu$m to $\sim$1.2 $\mu$m, as seen in Figure \ref{fig: LRD}. 
However, in both shorter and longer wavelengths, the shapes differ significantly. 
Specifically, in the longer wavelength range ($\lambda _{\rm obs} \gtrsim$ 2 $\mu$m), the emission from LRDs is significantly weaker than those of BPL DOGs. 
These two facts suggest that LRDs probably have similar dust extinction but week (or no) re-emission of hot dust heated by AGNs, relative to BPL DOGs. 
This result is consistent with SED properties of LRDs reported in previous studies (P\'{e}rez-Gonz\'{a}lez et al. \citeyear{Perez24}; Akins et al. \citeyear{Akins24}; Li et al. \citeyear{Li25}). 
Therefore, LRDs do not satisfy the optical-MIR red color criterion for DOGs. 
In other words, the DOG color criterion is not effective for selecting LRDs in the low-$z$ Universe.

\subsection{Selection bias for extremely red DOGs} \label{subsec:$Euclid$}

%% ========
%%  figure 
%% ========
\begin{figure}[t]
\includegraphics[width=8.5cm]{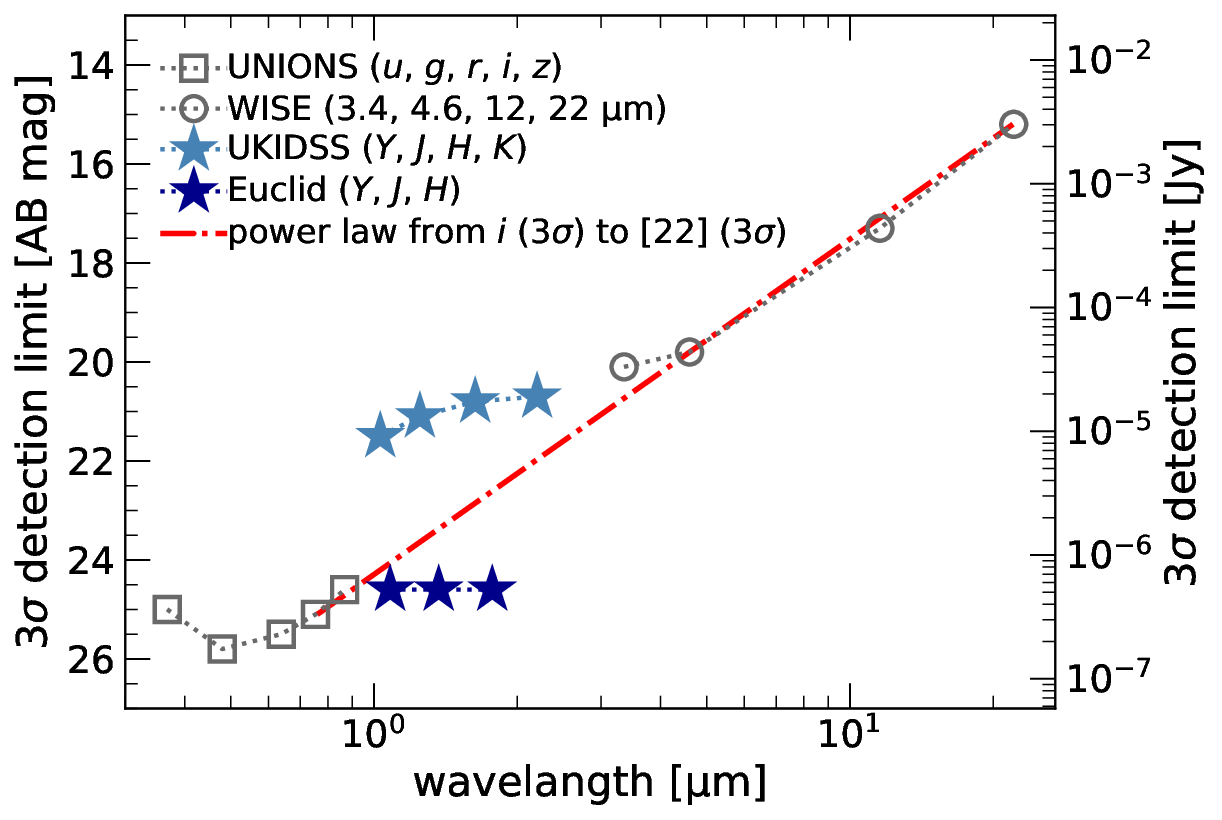}
\caption{
Comparison of the 3$\sigma$ detection limits of wide-field surveys (open squares: UNIONS, open circles: WISE, filled light-blue stars: UKIDSS, filled dark-blue stars: $Euclid$). 
Red dash-dotted line denotes a power law from the 3$\sigma$ detection limits of UNIONS $i$-band to WISE 22 $\mu$m band. 
Note that UNIONS $z$-band photometry is displayed in this figure, but we did not use $z$-band photometry in this work (see Section \ref{subsec:DOGs}). 
\label{fig: lims}}
\end{figure}
%% =====
%%  end
%% =====

In Section \ref{subsec:k-band}, we showed that a large fraction of BPL DOGs in our sample is caused by the shallow detection limit of UKIDSS. 
In addition to this issue, DOGs with an extremely red color ($i - [22] \gtrsim 7.5$) may be missing from our sample due to the shallow detection limit of UKIDSS. 
Figure \ref{fig: lims} shows the 3$\sigma$ detection limits of each survey. % and power law from those of UNIONS $i$-band to WISE 22 $\mu$m band. 
Suppose that a DOG with $i$-band and 22 $\mu$m magnitudes that are close to the detection limits of UNIONS and WISE, then such a DOG should show a very red color, i.e., $i - [22] \sim 10$ (like the red dash-dotted line shown in Figure \ref{fig: lims}). 
However, there are no such extremely red DOGs in our sample because such objects are not detected by UKIDSS, as seen in Figure \ref{fig: lims}. 
This is an extreme case, but a significant number of extremely red DOGs ($i - [22] \gtrsim 7.5$) are possibly missed in our sample because of our requirement of the K-band detection.

%% ========
%%  figure 
%% ========
\begin{figure}[t]
\includegraphics[width=8.5cm]{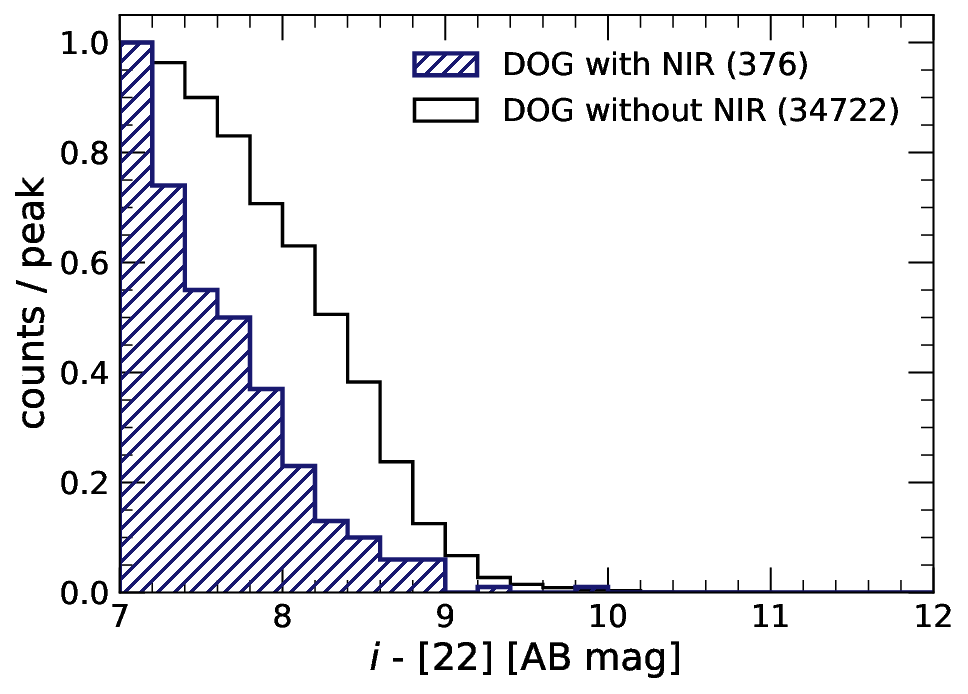}
\caption{
Comparison of $i - [22]$ color distribution of DOGs with NIR photometry (DOGs with NIR) and those of DOG without NIR photometry (DOG without NIR). 
power law from those of UNIONS $i$-band to WISE 22 $\mu$m.
The counts in each bin are normalized to the peak counts. 
The numbers in parentheses indicate the number of objects in each sample. 
\label{fig: wo_NIR}}
\end{figure}
%% =====
%%  end
%% =====

To investigate this possibility, we selected DOGs without requiring K-band detection and the $i - K$ red color criterion.
Specifically, we cross-matched the reliable UNIONS and WISE samples using a search radius of 3$^{\prime\prime}$ and adopted the definition of the DOG ($i - [22] \geq 7.0$). 
As a result, we found 34,722 DOGs (hereafter no-NIR DOGs) in the UNIONS2000 field ($\sim$ 2,000 deg$^2$). 
Note that a small fraction of these DOGs are mismatched due to the significant difference in the angular resolution between UNIONS and WISE (see Toba et al. \citeyear{Toba15}; Toba \& Nagao \citeyear{Toba16}; Noboriguchi et al. \citeyear{Nobo19}), given the fact that the pre-selection criterion of $i - K \geq 1.2$ is not adoped here. 
To investigate the fraction of mismatched sources, we counted the sources from our previous DOG sample (i.e., with NIR photometry) that are included in the no-NIR DOGs. 
In other words, we searched for WISE sources that have the same UNIONS counterparts, regardless of our the optical-NIR color cut.
As a result, 356/376 ($\sim$ 95\%) remained, meaning that $\sim$ 5\% of the sources were mismatched. 
We regard this fraction is negligibly low for our discussion here.
Figure \ref{fig: wo_NIR} shows that the comparison of $i - [22]$ color of no-NIR DOGs and those of our previous DOG sample (i.e., with NIR photometry). 
The median of $i - [22]$ for no-NIR DOGs and our previous DOG sample are 7.7 and 7.4, respectively. 
Figure \ref{fig: wo_NIR} suggests that a large fraction of DOGs with extremely red color ($7.5 \lesssim i - [22] \lesssim 10.0$) are actually missed in our sample due to the shallow detection limit of UKIDSS. 

However, this issue will soon be resolved by $Euclid$ (and subsequent Roman space telescope), as it is 3--4 magnitudes more sensitive than UKIDSS in NIR (see Figure \ref{fig: lims}). 
By combining UNIONS and $Euclid$ data, we will be able to construct statistically large samples of DOGs with a less-biased selection.
Assuming the same number density as no-NIR DOGs (i.e., $\sim 2 \times 10^1$ deg$^{-2}$), we expect to identify $\sim$ 90,000 DOGs using the full datasets of UNIONS and $Euclid$ (over $\sim$ 5,000 deg$^2$).
Additionally, this sample will allow us to study a significant number of DOGs with extremely red colors, which have been incompletely studied due to their low number density.

\section{Conclusion} \label{sec:conlusion}

We identified 382 DOGs in the overlap region of the UNIONS and UKIDSS survey areas ($\sim$ 170 deg$^2$) and classified the vast majority of them (376 DOGs) into bump (SF dominated) DOGs and PL (AGN dominated) DOGs.
During the SED analysis, we found that roughly half (120/244) of the PL DOGs show ``broken'' power-law SEDs. 
The properties of these ``broken-PL (BPL) DOGs'' are as follows:

\begin{enumerate}
\item The MIR slope of BPL DOGs indicates the presence of hot dust emission heated by an AGN, but their optical slope are more redder than those of normal-PL (NPL) DOGs (Figure \ref{fig: slope}), suggesting that BPL DOGs are more heavily obscured AGNs (see Section \ref{subsec:slope}, and \ref{subsec:heavy}).
\item The differences in the SED shapes between NPL and BPL DOGs can potentially be explained by either the presence of a torus extinction (viewing angle scenario; Section \ref{subsubsec:viewing}) and/or the amount of host extinction (transition phase scenario; Section \ref{subsubsec:transition}).
\item The large fraction of BPL DOGs in our sample is likely due to a combination of their higher NIR flux and the shallow detection limit in NIR. 
Their higher NIR flux is potentially caused by the rest-frame 1.6 $\mu$m bump, which possibly supports the transition phase scenario (Section \ref{subsec:k-band}). 
\item The SED shape of BPL DOGs matches well with the torus model and heavily reddened type-1 quasars (Section \ref{subsec:heavy}), whereas it bears some resemblance to that of Hot DOGs and LRDs but differs in detail (see Section \ref{subsec:dusty}).
\end{enumerate}

\section*{acknowledgments}
%\begin{acknowledgments}
The authors gratefully acknowledge the anonymous referee for a careful reading of the manuscript and very helpful comments.

% UNIONS
We are honored and grateful for the opportunity of observing the Universe from Maunakea and Haleakala, which both have cultural, historical and natural significance in Hawaii. 
This work is based on data obtained as part of the Canada-France Imaging Survey, a CFHT large program of the National Research Council of Canada and the French Centre National de la Recherche Scientifique. 
Based on observations obtained with MegaPrime/MegaCam, a joint project of CFHT and CEA Saclay, at the Canada-France-Hawaii Telescope (CFHT) which is operated by the National Research Council (NRC) of Canada, the Institut National des Science de l’Univers (INSU) of the Centre National de la Recherche Scientifique (CNRS) of France, and the University of Hawaii. 
This research used the facilities of the Canadian Astronomy Data Centre operated by the National Research Council of Canada with the support of the Canadian Space Agency. 
This research is based in part on data collected at Subaru Telescope, which is operated by the National Astronomical Observatory of Japan. 
Pan-STARRS is a project of the Institute for Astronomy of the University of Hawaii, and is supported by the NASA SSO Near Earth Observation Program under grants 80NSSC18K0971, NNX14AM74G, NNX12AR65G, NNX13AQ47G, NNX08AR22G, 80NSSC21K1572 and by the State of Hawaii.

%UKIDSS
This work is based in part on data obtained as part of the UKIRT Infrared Deep Sky Survey.

%WISE
This publication makes use of data products from the Wide-field Infrared Survey Explorer, which is a joint project of the University of California, Los Angeles, and the Jet Propulsion Laboratory/California Institute of Technology, funded by the National Aeronautics and Space Administration.
%\end{acknowledgments}

This work was financially supported by JSPS KAKENHI grants: Nos. 20H01949 (T.N.), 23H01215 (T.N.), 23K22537 (Y.T.) and 24K00684 (M.O.).
H. H. is supported by a DFG Heisenberg grant (Hi 1495/5-1), the DFG Collaborative Research Center SFB1491, an ERC Consolidator Grant (No. 770935), and the DLR project 50QE2305.
Y.Z. is supported by JSPS Research Fellowship for Young Scientists.
This work was supported by JSPS Core-to-Core Program (grant number: 
JPJSCCA20210003).

%appendix

\bibliography{bib_Y+25}{}
\bibliographystyle{aasjournal}

%% the collaboration and author truncation commands are used. It has to


\end{document}